\documentclass[english, reprint,prl, twocolumn, superscriptaddress,showpacs,amsmath,amssymb,nofootinbib]{revtex4-1}
\usepackage{lmodern}
\usepackage[T1]{fontenc}
\usepackage[latin9]{inputenc}
\usepackage[pdftex]{color}
\usepackage{amstext}
\usepackage{amssymb}
\usepackage[pdftex]{graphicx}
\usepackage{hyperref}

\makeatletter

\pdfpageheight\paperheight
\pdfpagewidth\paperwidth

\newcommand{\lyxmathsym}[1]{\ifmmode\begingroup\def\b@ld{bold}
  \text{\ifx\math@version\b@ld\bfseries\fi#1}\endgroup\else#1\fi}

\@ifundefined{textcolor}{}
{%
 \definecolor{BLACK}{gray}{0}
 \definecolor{WHITE}{gray}{1}
 \definecolor{RED}{rgb}{1,0,0}
 \definecolor{GREEN}{rgb}{0,1,0}
 \definecolor{BLUE}{rgb}{0,0,1}
 \definecolor{CYAN}{cmyk}{1,0,0,0}
 \definecolor{MAGENTA}{cmyk}{0,1,0,0}
 \definecolor{YELLOW}{cmyk}{0,0,1,0}
}

\usepackage{upgreek}
\usepackage{babel}

\newcommand{\SFOO}{Sr$_2$FeOsO$_6$}
\newcommand{\sfoo}{Sr$_2$FeOsO$_6$}
\newcommand{\BYOO}{Ba$_2$YOsO$_6$}
\newcommand{\BYRO}{Ba$_2$YRuO$_6$}

\newcommand{\SSOO}{Sr$_2$ScOsO$_6$}
\newcommand{\LNRO}{La$_2$NaRuO$_6$}

\newcommand{\CLOO}{Ca$_3$LiOsO$_6$}
\newcommand{\SCOO}{Sr$_2$CrOsO$_6$}

\newcommand {\TN} {$T_{\mathrm{N}}$}
\newcommand {\TNo}{$T_{\mathrm{N}1}$}
\newcommand {\TNt} {$T_{\mathrm{N}2}$}

\newcommand {\etal} {\emph{et al.}}

\makeatother

\begin{document}

\begin{abstract}
\sfoo{} is an insulating double perovskite compound which undergoes antiferromagnetic transitions at 140 K ($T_{N1}$) and 67 K ($T_{N2}$).  To study the underlying electronic and magnetic interactions giving rise to this behavior we have performed inelastic neutron scattering (INS) and resonant inelastic x-ray scattering (RIXS) experiments on polycrystalline samples of \sfoo{}.  The INS data reveal that the spectrum of spin excitations remains ungapped below T$_{N1}$, however below T$_{N2}$ a gap of 6.8 meV develops. The RIXS data reveals splitting of the T$_{2g}$ multiplet consistent with that seen in other 5d$^3$ osmium based double perovskites.  Together these results suggest that spin-orbit coupling is important for ground state selection in 3d-5d$^3$ double perovskite materials.\footnote{This manuscript has been authored by UT-Battelle, LLC under Contract No. DE-AC05-00OR22725 with the U.S. Department of Energy.  The United States Government retains and the publisher, by accepting the article for publication, acknowledges that the United States Government retains a non-exclusive, paid-up, irrevocable, world-wide license to publish or reproduce the published form of this manuscript, or allow others to do so, for United States Government purposes.  The Department of Energy will provide public access to these results of federally sponsored research in accordance with the DOE Public Access Plan (http://energy.gov/downloads/doe-public-access-plan).}
\end{abstract}

\title{Origin of magnetic excitation gap in double perovskite \sfoo{}}

\author{A. E. Taylor }

\address{Neutron Scattering Division, Oak Ridge National Laboratory,
Oak Ridge, Tennessee 37831, USA}

\author{R. Morrow}

\address{Department of Chemistry, The Ohio State University, Columbus, Ohio
43210-1185, USA}

\address{Leibniz Institute for Solid State and Materials Research Dresden IFW, Dresden D-01069, Germany}

\author{M. D. Lumsden}

\address{Neutron Scattering Division, Oak Ridge National Laboratory,
Oak Ridge, Tennessee 37831, USA}

\author{S. Calder}

\address{Neutron Scattering Division, Oak Ridge National Laboratory,
Oak Ridge, Tennessee 37831, USA}

\author{M. H. Upton}

\address{Advanced Photon Source, Argonne National Laboratory, Argonne, Illinois
60439, USA}

\author{A. I. Kolesnikov}

\affiliation{Neutron Scattering Division, Oak Ridge National Laboratory,
Oak Ridge, Tennessee 37831, USA}

\author{M. B. Stone}

\address{Neutron Scattering Division, Oak Ridge National Laboratory,
Oak Ridge, Tennessee 37831, USA}

\author{R. S. Fishman }
\affiliation{Materials Science and Technology Division, Oak Ridge National Laboratory,
Oak Ridge, Tennessee 37831, USA}

\author{A. Paramekanti}
\affiliation{Department of Physics, University of Toronto, Toronto, Ontario M5S 1A7, Canada}
\affiliation{Canadian Institute for Advanced Research, Toronto, Ontario, M5G 1Z8, Canada}

\author{P. M. Woodward}

\address{Department of Chemistry, The Ohio State University, Columbus, Ohio
43210-1185, USA}

\author{A. D. Christianson}
\email{christiansad@ornl.gov}

\affiliation{Materials Science and Technology Division, Oak Ridge National Laboratory,
Oak Ridge, Tennessee 37831, USA}

\address{Neutron Scattering Division, Oak Ridge National Laboratory,
Oak Ridge, Tennessee 37831, USA}


\maketitle
%


A strong spin-orbit interaction is inherent to 4$d$ and 5$d$ ions,
and when this is manifest in the collective properties of materials
via spin-orbit coupling (SOC) it fosters a host of unconventional
phases. For example, SOC is responsible for
the $J_{\mathrm{eff}}=1/2$ electronic ground state which leads
to a Mott insulating phase in $5d^{5}$ Sr$_{2}$IrO$_{4}$~\cite{kim_phase-sensitive_2009},
and Kitaev quantum-spin-liquid-like behavior in $4d^{5}$ RuCl$_{3}$~\cite{banerjee_proximate_2016}. Beyond this $J_{\mathrm{eff}}=1/2$ paradigm,
however, the influence of the spin-orbit interaction on the electronic
ground state and emergent properties in $4d$ and $5d$ transition
metal oxides (TMOs) has been poorly understood.

Recently, a SOC-controlled $J=3/2$ ground state was discovered in
$5d^{3}$ TMOs \BYOO{} and \CLOO{} \cite{alice_2017}, in contrast with expectations
of an orbitally quenched $S=3/2$ singlet. Ref. \onlinecite{alice_2017} revealed a SOC-induced
splitting of the $t_{2g}^{3}$ manifold via resonant inelastic x-ray
scattering (RIXS) measurements, which is driven by
strong Os-O hybridization. This confirmed the presence of an unquenched
orbital moment in the $d^{3}$ ion ground state, and placed these
$5d^{3}$ materials in the intermediate coupling regime, between
$LS$ and $jj$ coupling limits. There is therefore immediate interest
in exploring the impact of this $J=3/2$ ground state on the emergent
properties in $d^{3}$ TMOs. 

\begin{figure}[t]
\includegraphics[width=0.9\columnwidth]{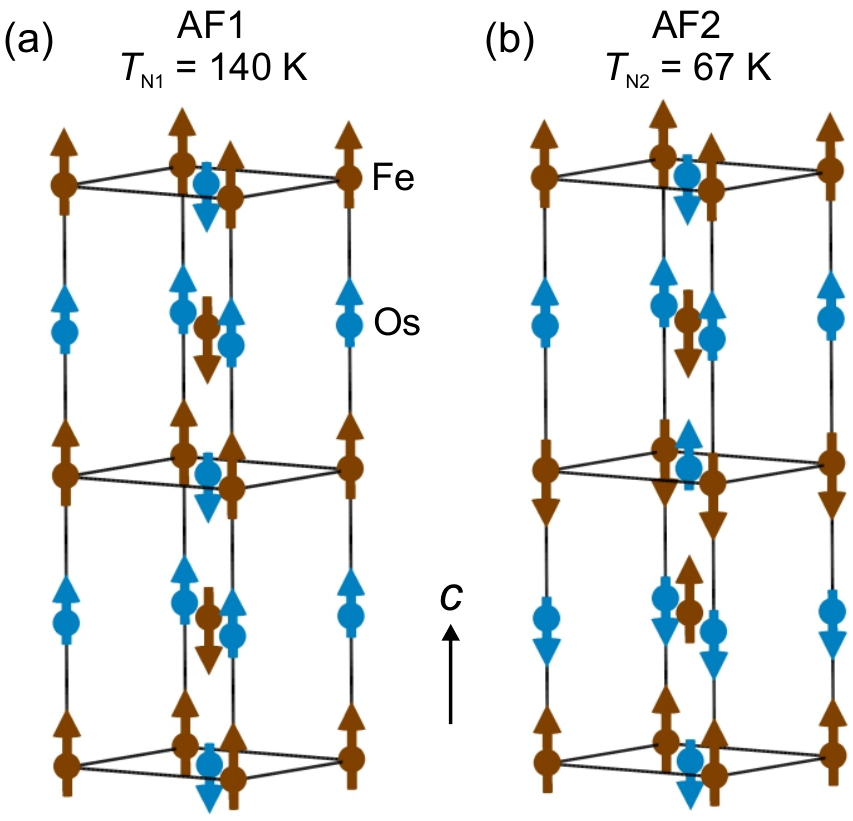}\protect\caption{\label{fig:magstruct}
Magnetic structure of \SFOO{} determined by Refs. \onlinecite{paul_lattice_2013,adler_magnetic_2013}. (a) AF1 phase
which is dominant between 140\,K and 67\,K and (b) AF2 phase which
emerges at 67\,K.  The Os (Fe) atoms and direction of the ordered magnetic moment are indicated by the blue (brown) circles and arrows.  Two crystallographic unit cells are shown so that the changes between magnetic phases are apparent.  }
\end{figure}

In the cubic double perovskite \BYOO{} the direct influence of the $J=3/2$
SOC has been observed via a spin-gap in the magnetic excitation spectrum with inelastic neutron scattering~\cite{kermarrec_frustrated_2015}.  More broadly, spin gaps have been observed in many single magnetic ion containing 4d$^3$ and 5d$^3$ double perovskites  and related materials \cite{carlo_spin_2013,aczel_exotic_2014,kermarrec_frustrated_2015, disseler_2016, taylor_spin-orbit_2016, calder_2017}.
These results indicate that SOC directly influences the magnetic ground state
in otherwise frustrated systems, which provides scope for
control of the physical properties of such materials via SOC.  However,
for practical functional materials, the role which SOC
plays in significantly non-cubic materials, and in systems with more
complex interactions e.g. mixed 3$d$-5$d$ systems, are open questions. Here, we investigate these issues in the 3$d$-5$d$ material
\SFOO{}.

\begin{figure}
\includegraphics[clip,width=0.98\columnwidth]{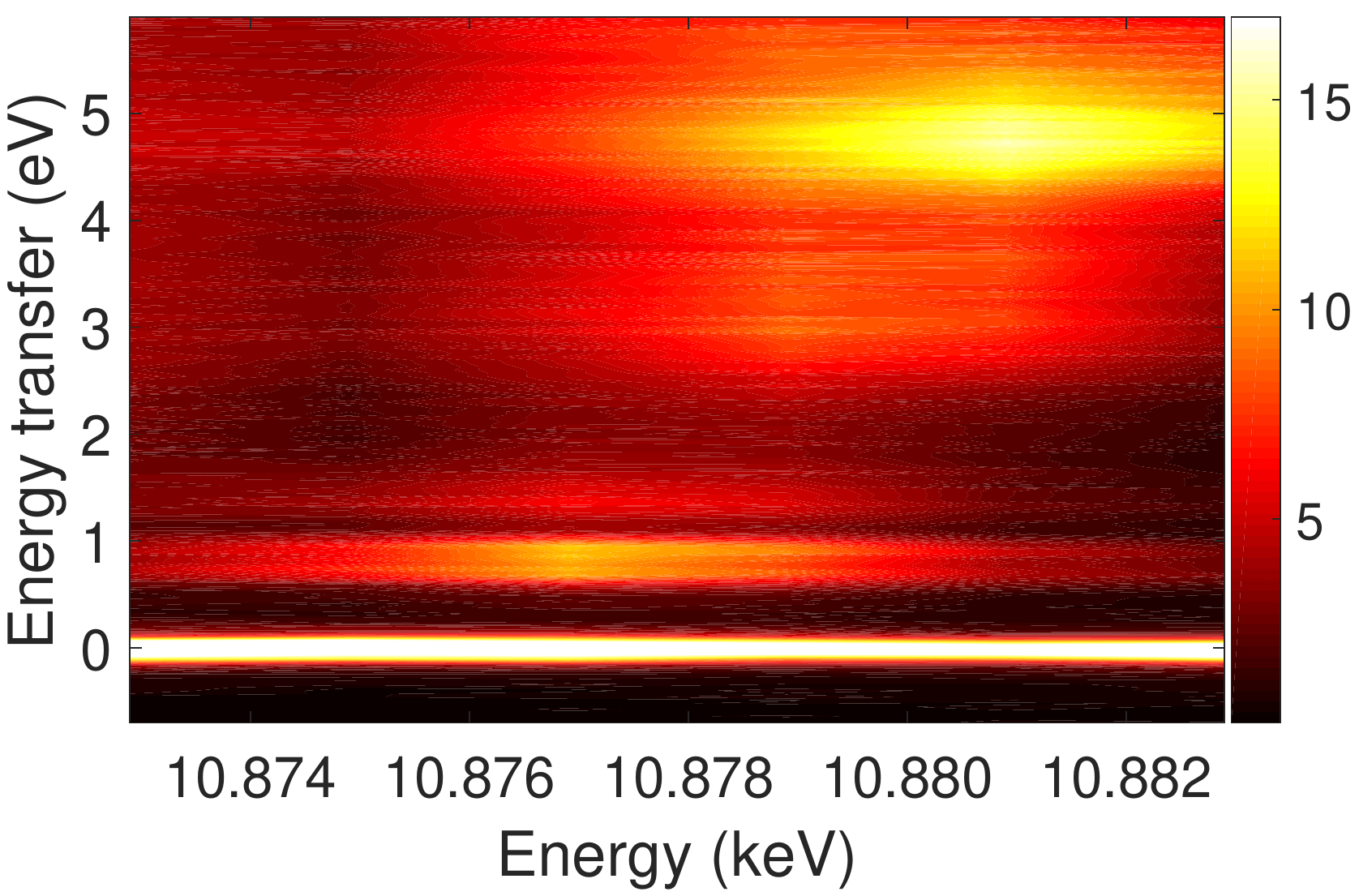}

\caption{RIXS data showing the incident energy dependence of electronic excitations
in \SFOO{}. Measurements were performed at 6 K. The color bar on the right indicates intensity in arbitrary units.}
\label{RIXS_cmap}
\end{figure}

\SFOO{} has attracted a great deal of attention due to highly tunable
magnetic behaviour~\cite{adler_magnetic_2013,feng_high-temperature_2014,morrow_probing_2014,paul_synthesis_2013,paul_lattice_2013,veiga_fragility_2015,williams_muon-spin_2015}, and because it is an unusual anomaly in
the important series of compounds Sr$_{2}$\emph{BB}$^{\prime}$O$_{6}$,
where \emph{B} and $B^{\prime}$ are 3\emph{d} and 4\emph{d} or 5\emph{d}
magnetic transition metal ions, respectively. These materials are
well-known for their potential spintronics applications, and they
generally present above room-temperature ferrimagnetism and evolve
from half-metallic to insulating states as $T_{\mathrm{C}}$ increases:
Sr$_{2}$FeReO$_{6}$ $T_{\mathrm{{C}}}=401\,$K, Sr$_{2}$FeMoO$_{6}$
$T_{\mathrm{{C}}}=420\,$K, Sr$_{2}$CrWO$_{6}$ $T_{\mathrm{{C}}}=450\,$K,
Sr$_{2}$CrReO$_{6}$ $T_{\mathrm{{C}}}=625\,$K, and \SCOO{} $T_{\mathrm{C}}=725\,$K~\cite{vasala_a2bbo6_2015}.
However, \SFOO{} with Os$^{5+}$ (5d$^3$, $S=\frac{3}{2}$) and Fe$^{3+}$ (3d$^5$, $S=\frac{5}{2}$) is an insulating antiferromagnet which undergoes
two magnetic transitions at much lower temperatures, $T_{\mathrm{1}}=140\,$K
and $T_{\mathrm{2}}=67\,$K with Fe and Os ordering in both of the magentic phases(Fig.~\ref{fig:magstruct}) \cite{paul_lattice_2013}.

The presence of two transitions in \SFOO{} suggests competing magnetic interactions. The competition in \SFOO{} is further demonstrated by the ease with
which it can be tuned to other magnetic ground states, either by isoelectronic
doping on the \emph{A} site~\cite{morrow_probing_2014}, or via hydrostatic
pressure~\cite{veiga_fragility_2015}, which opens a route to strain-controlled
epitaxial films as functional devices~\cite{veiga_fragility_2015}. Multiple first
principles calculations have attempted to identify the interactions
controlling this system, yet have produced disparate results with
predictions including semiconductor behavior~\cite{wang_first_2011,wang_first-principle_2014},
orbital order~\cite{paul_lattice_2013}, dominant Fe-O-Os superexchange
interactions~\cite{hou_lattice-distortion_2015}, or dominant Os-O-O-Os
extended superexchange~\cite{kanungo_textitab_2014}. None of these works considered the potential role of the recently discovered $J=3/2$ ground state possible for the Os$^{5+}$ ions\cite{alice_2017}.

Here we experimentally probe \SFOO{} via inelastic neutron scattering (INS) and resonant inelastic x-ray scattering (RIXS). We find
that despite the tetragonal distortion, SOC induced splitting of the
Os$^{5+}$ $t_{2g}$ levels is observed, indicative of a $J=3/2$ ground state. This provides a route to
strong entry of SOC in this material - a factor which has not previously
been explored - and this conjecture is confirmed, as we reveal a spin-gap
in the magnetic excitation spectrum via INS. Previously no such feature
has been identified in a 3$d$-5$d$ TMO, only in purely 4$d$ or
5$d$ materials. Unexpectedly, this SOC-induced gap only emerges below
the second magnetic ordering temperature $T_{\mathrm{2}}=67\,$K.
This suggests that SOC is likely intimately linked to the selection
of the ground state in \SFOO{} via SOC-induced anisotropy, similar
to \BYOO{} and \SSOO{} \cite{alice_2017,taylor_spin-orbit_2016}. These considerations should also apply to other 3d-5d$^3$ combinations such as the high ordering temperature Sr$_2$CrOsO$_6$ \cite{krockenberger_sr2croso6:_2007,morrow_2016d}.


\begin{figure}
\includegraphics[clip,width=0.98\columnwidth]{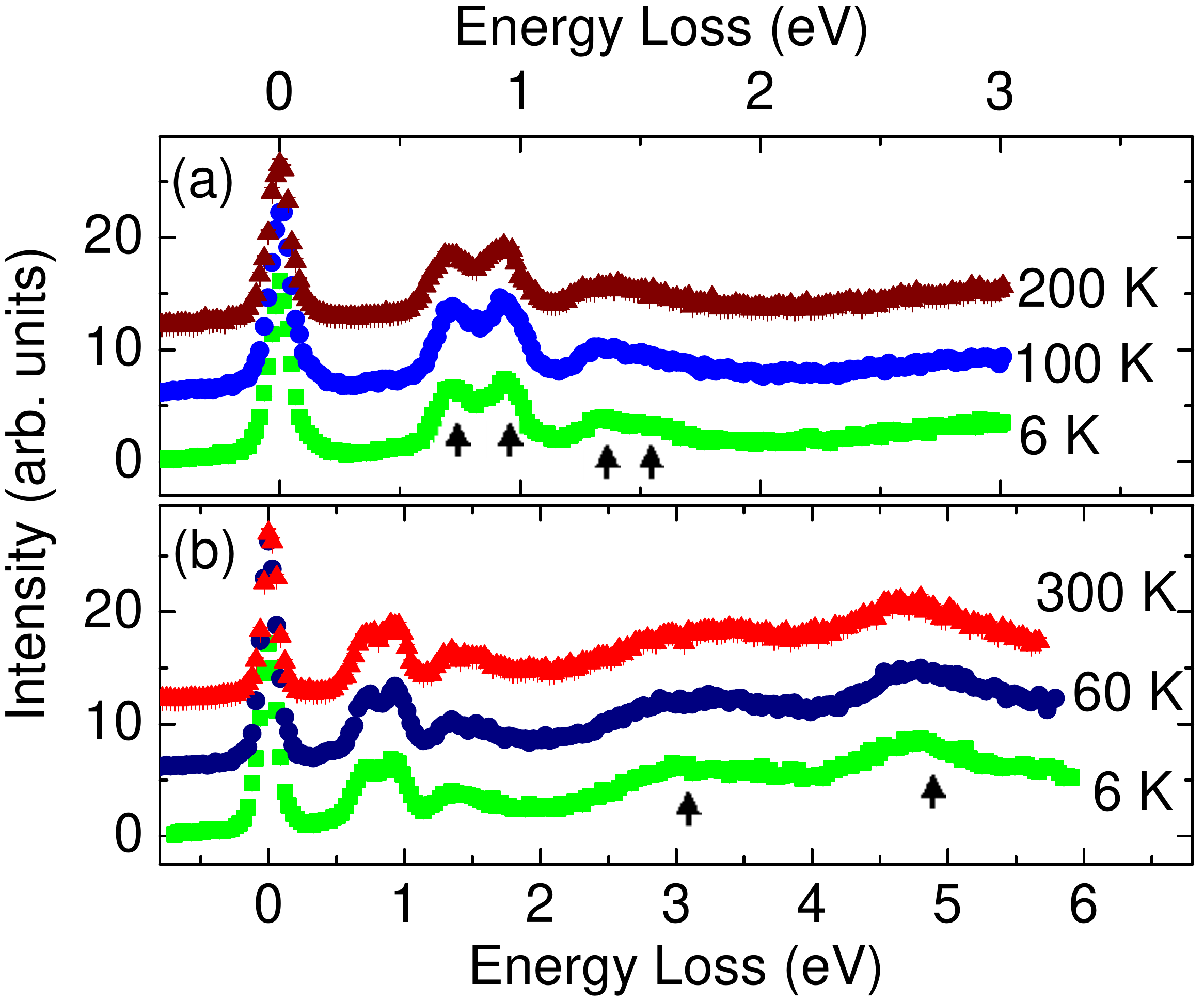}

\protect\caption{\label{RIXS} RIXS data from \SFOO{}. (a) High resolution data with $E_i$ = 10.877 keV showing excitations (indicated by arrows) within the t$_{2g}$ multiplet. (b) Low resolution RIXS data with $E_i$ = 10.879 keV.  The arrows indicate the positions of excitations from the $t_{2g}$ to the $e_g$ multiplets. In (a) and (b) an offset of 6 between data sets has been added for clarity.}
\end{figure}

The 12.8\,g polycrystalline \SFOO{} sample was synthesized by combining
stoichiometric quantities of SrO$_{2}$, Os, OsO$_{2}$ and Fe$_{2}$O$_{3}$.
Ground mixtures were contained in alumina tubes and sealed in evacuated
silica vessels for heatings of 48\,hours at 1000$\,\textdegree$C.
This was followed by a regrinding and identical reheating. Laboratory x-ray diffraction measurements were performed to characterize the structural order of the samples studied here.  Additional characterization of the structural and magnetic order was provided by neutron powder diffraction measurements performed with HB-2A at
the High Flux Isotope Reactor at Oak Ridge National Laboratory (ORNL).   The results of the x-ray and neutron diffraction measurements are given in the supplemental material\cite{supp}.
 
Inelastic neutron scattering measurements were performed with the SEQUOIA
chopper spectrometer \cite{stone_seq} at the Spallation Neutron Source at ORNL. The
sample was sealed in a flat plate Al cell with 2\,mm thickness in
order to minimize the effects of absorption. This cell and an identical
empty cell were measured in a closed-cycle refrigerator, accessing
temperatures between 5\,K and 170\,K. Incident neutron energies
($E_{\mathrm{i}}$s) of 20 and 60\,meV, with fermi chopper frequencies
of 120 and 180\,Hz respectively, were used. Empty-cell background
data has been subtracted from all datasets presented.

RIXS spectra were collected on a small portion of the sample on Sector
27 at the Advanced Photon Source (APS) using the MERIX instrumentation~\cite{gog_momentum-resolved_2009}.
The sample temperature was controlled between 6$\,$K and 300$\,$K
in a closed-cycle refrigerator. Primary diamond(1 1 1) and secondary
Si(4 0 0) monochromators were used to access the Os L$_{3}$-edge
E$_i$, with a diced Si(4 6 6) analyzer to discriminate the
scattered beam energy yielding a energy resolution of 125 meV FWHM.  Some scans were also collected with a channel cut Si (4 6 6) analyzer yielding an energy resolution of 55 meV FWHM.  A MYTHEN strip detector was used, and experiments
were performed in horizontal geometry with $2\mathrm{\theta}=90^{\circ}$.
Data are normalized to the incident beam intensity via an ion chamber
monitor.


Figure \ref{RIXS_cmap} displays the excitation spectrum of \SFOO{} measured with RIXS as a function of $E_i$.  As in the case of  other osmium-based TMOs\cite{calder_spin-orbit-driven_2016,calder_2017c,alice_2017}, the relative energies of the inelastic features along with the dependence of the inelastic features on E$_i$ allows for the identification of intra $t_{2g}$ processes (maximum intensity near 10.877 keV) and $t_{2g}-e_g$ processes (maximum intensity near 10.881 keV).  

Figure~\ref{RIXS} shows the RIXS spectra from \SFOO{} measured
at selected temperatures in high resolution (a) and low resolution configurations (b). SOC-induced splitting
of the $t_{2g}$ character excited state is apparent at all temperatures
in the peaks centered around \textasciitilde{}0.75$\,$eV. Peak splitting
around \textasciitilde{}1.5\,eV is not resolved, but the width which is significantly broader than instrumental resolution and the asymmetric shape of the signal
indicate that two peaks are likely present, as found in \BYOO{} and \CLOO{}\cite{alice_2017}. At present it is not clear why the two peaks are not well resolved.  Possibilities include the overall tetragonal symmetry and the antisite mixing with 0.856(4) Fe (Os) and 0.144(4) Os (Fe) occupancy on the \emph{B} (\emph{B}$^\prime$) site \cite{supp} which is within the typical range for this material~\cite{paul_synthesis_2013,paul_lattice_2013,morrow_probing_2014}.

\begin{figure}
\includegraphics[clip,width=1\columnwidth]{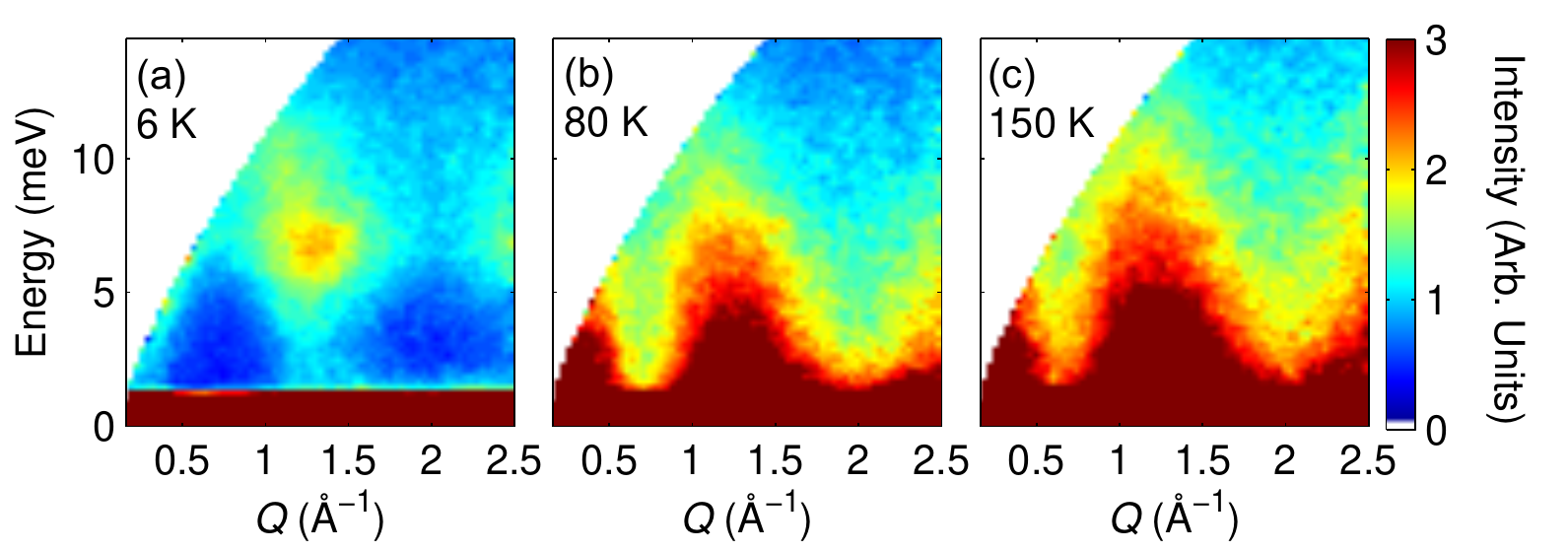}
\caption{Neutron scattering intensity maps showing the evolution
of the scattering from (a) T = 6 K (T < \TNt{}), (b) T = 80 K (\TNt{} < T < \TNo{}), (c) T = 150 K (T > \TNo{}).  The incident energy was 20 meV. A
gap in the magnetic excitations emerges below \TNt{} as seen in (a). }
\label{Slices}
\end{figure}

The high resolution RIXS data were fit with a Lorentizian peak shape, giving excited state energies of 0.717(3), 0.936(4), 1.34(7) and 1.5(2)\,eV
at 6\,K. This compares with 0.745(7), 0.971(7), 1.447(9) and 1.68(1)eV
from Ref. \onlinecite{alice_2017} for  \BYOO{}. As was done in Ref. \onlinecite{alice_2017}, these data can be fit with an intermediate coupling model (under the assumption of cubic crystal field splitting) to extract values of the spin orbit coupling,~$\zeta_{SOC}$ , and Hund's coupling, $J_h$.  To prevent proliferation of fitting parameters we have fixed the Racah parameter B to the value of 0.0405 eV which was determined by Ref. \onlinecite{dorain_optical_1966} for 5d$^3$ Re$^{4+}$.  Fixing B to other values over a relatively broad range results in similar values of the Hunds's coupling. The model parameters are:  The Racah parameters B (fixed), C, the crystal field splitting 
$10Dq$ (fixed) , and the spin-orbit coupling $\zeta_{SOC}$. The results of the fits are insensitive to values of $10Dq$ from 3 to 4.8 eV.  The resulting parameters of the fits are: C = 0.21(1) eV and $\zeta_{SOC}$ = 0.33(7) eV.  $J_h$ = 3B+C=0.27(1) eV.  The values are comparable to those found for \BYOO{} and \CLOO\cite{alice_2017}, other related 5d containing double perovskite systems\cite{other_hunds1,other_hunds2} and for NaOsO$_3$ \cite{James_2018}.

\begin{figure}
\includegraphics[clip,width=0.9\columnwidth]{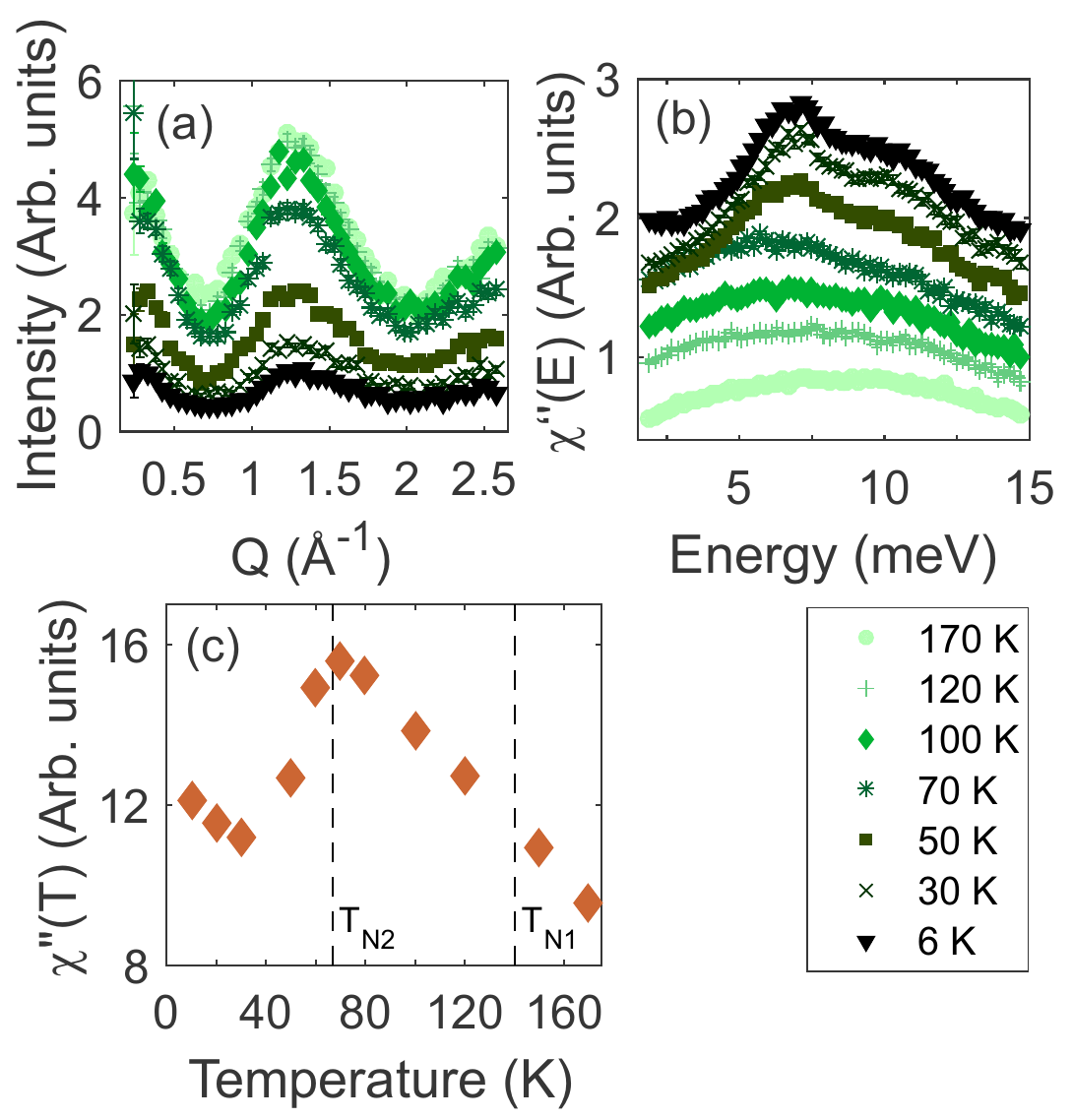}

\caption{(a) Measured intensity of neutron scattering data averaged over 2--3\,meV
at temperatures as indicated in the legend in the lower right. 
(b) Constant-wave vector cuts averaged over 0.8--1.8\,$\AA^{-1}$which
have been corrected for the Bose population factor, $1/(1-\exp(-E/k_{\mathrm{B}}T))$.
Successive cuts have been offset by 0.2.  (c) Data averaged over
2--3\,meV and 0.8--1.8\,$\AA^{-1}$ and corrected
for the Bose factor as a function of temperature, demonstrating temperature
dependence of the scattering within the gap. All panels show data collected
with $E_{\mathrm{i}}=20\,$meV. In panels (b) and (c) the errorbars
are smaller than the symbols.}
\label{INS_cuts}
\end{figure}

A distinction between the RIXS data for \SFOO{}  and other $5d^3$ systems is that there appears to be two excitations between the $t_{2g}$ and the $e_g$ multiplets.  These are found to be at 3.05(3) and 4.85(6) eV (indicated by arrows in Fig. \ref{RIXS}(b)). The overall tetragonal symmetry of \SFOO{} is likely not the reason for the appearance of the two features.  At the lowest temperatures the oxygen octahedra surrounding the osmium ions are nearly cubic \cite{paul_synthesis_2013,morrow_probing_2014} and both features are observed in the RIXS spectra at all temperatures measured with minimal variation.  Another possibility is that the antisite mixing between Fe and Os provides two environments for Os resulting in the two peak structure. A more interesting possibility is that the Hund's coupling on Fe and strong Os-O and O-Fe hybridization for $e_g$ orbital results in a large splitting for spin flip and spin parallel excitations from the $t_{2g}$ to the $e_g$. Such a scenario might, however, lead to a temperature dependent peak splitting depending on the evolution of 
magnetic correlations on Fe sites.

An overview of the measured INS spectra from 6, 80 and 150\,K is
shown in Fig.~\ref{Slices}. Whilst there is no significant change
in the inelastic spectrum upon crossing \TNo{}, below \TNt{} there
is a pronounced change in the excitation spectrum, as a gap opens
and the intensity is concentrated at higher energies, see 6\,K data.  
This behavior is reminiscent of the observed gap development below
\TN{} in the previously measured single-magnetic-ion 4\emph{d}$^{3}$
and 5\emph{d}$^{3}$ double perovskites~\cite{aczel_exotic_2014,carlo_spin_2013,kermarrec_frustrated_2015,taylor_spin-orbit_2016}.  Note that at the lowest temperatures where the spin gap is the strongest the octahedra surrounding the osmium ions are more symmetric\cite{paul_synthesis_2013,morrow_probing_2014} than at higher temperatures indicating that a structural distortion at \TNt{} is not likely the origin of the observed spin gap.

The detailed $(Q,E)$-space temperature dependence
of the scattering is presented in Fig.~\ref{INS_cuts}. Constant-energy
cuts averaged over $2<E<3\,$meV, i.e. within the gap, are shown
in Fig.~\ref{INS_cuts}(a). Intensity is observed around wave vector
$Q\approx0.4\,\mathrm{\AA^{-1}}$, which we attribute to scattering near the magnetic
wave vector $Q_{\mathrm{AF2}}=|(0\,0\,\frac{{1}}{2})|=0.39$ $\AA^{-1}$.
$Q_{AF2}=(0\,0\,\frac{{1}}{2})$ is not an observed
magnetic Bragg reflection because the magnetic moments lie along the
\emph{c} axis and only moments perpendicular to ${Q}$ give neutron scattering intensity. Transverse magnetic fluctuations
emerging from $(0\,0\,\frac{{1}}{2})$, however, have a moment perpendicular
to ${Q}$ and can therefore be observed, as in Figs.~\ref{Slices} and~\ref{INS_cuts}(a). Therefore the scattering at this purely AF2 wavevector
are associated with interactions responsible for the AF2 magnetic order.
The $Q\approx1.3\,\mathrm{\AA^{-1}}$ centered signal in Figs. \ref{Slices} and ~\ref{INS_cuts}(a)
 is a combination of scattering from $\bm{Q}_{\mathrm{AF1}}=(1\,0\,0)$,
and $\bm{Q}_{\mathrm{AF2}}=(0\,0\,\frac{{3}}{2})$ and $(1\,0\,\frac{{1}}{2})$
magnetic wave vectors, which cannot be resolved in this measurement.
Inspecting Fig.~\ref{INS_cuts}(a) it is clear that the fluctuations
do not go to zero within the gap at 6$\,$K, which is the result of
the presence of a significant fraction (31\%) of gapless phase 1 remaining
at this temperature (see Fig. S2 and associated discussion\cite{supp} and Refs. \onlinecite{paul_lattice_2013,adler_magnetic_2013}).

To track the relative strength of the fluctuations with temperature,
the data in Fig.~\ref{INS_cuts}(b) and (c) have been corrected for
the Bose thermal population factor, giving the results in terms of
the dynamic susceptibility $\chi^{\prime\prime}(Q,E)$. The constant-wave vector
cuts, averaged over 0.8--1.8$\,\mathrm{\AA}^{-1}$, in Fig.~\ref{INS_cuts}(b)
show that there is a significant build up of spectral weight in the
range $\sim5$--13\, meV below \TNt{}. There is little change in
the energy dependence of the scattering at the first transition, $T_{\mathrm{N1}}=140\,$K.
The onset of the gapped magnetic scattering intensity at low temperature
appears to be at $E\approx5\,$meV, and the peak of the intensity
is at 6.8(1)\,meV, determined by fitting two Gaussians on a flat
background to the 6\,K cut shown in Fig.~\ref{INS_cuts}(b). The
second peak in intensity is determined to be 10.5(1)$\,$meV\textcolor{black}.
The observation of two peaks within the band is consistent with the
effects of powder averaging the magnetic excitation signals originating
from the inequivalent directions present in the tetragonal crystal
structure, as seen in other significantly distorted double perovskites~\cite{aczel_exotic_2014}.
We compare the peak of the lower band, $\Delta=6.8(1)$\,meV, to
previous observations of the gap in 4\emph{d$^{3}$} and 5\emph{d$^{3}$}
double perovskites, as these have followed the convention of using the center of
the acoustic band as an estimate of the value of the gap. In \SSOO{},
\BYOO{}, \BYRO{} and \LNRO{} the determined values are $\Delta=19(2)\,$meV,
$\Delta=18(2)\,$meV, $\Delta\approx5\,$meV and $\Delta\approx2.75\,$meV,
respectively~\cite{kermarrec_frustrated_2015,carlo_spin_2013,aczel_exotic_2014}.
Notably, the energy scale of the gap in \SFOO{} is significantly lower than
Os$^{5+}$ counterparts \SSOO{} and \BYOO{}, but still larger than
the Ru$^{5+}$ examples, in which spin-orbit effects are expected
to be reduced. 

Another method of estimating the gap was used by Kermarrec \etal{}~\cite{kermarrec_frustrated_2015}
for \BYOO{}, in which $\chi^{\prime\prime}(T)$ for $E<\Delta$ is
compared to that expected for a thermally activated excitation, i.e.,
$\chi^{\prime\prime}(T<T_{\mathrm{N2}})\propto\exp(-\Delta/k_{\mathrm{B}}T)$.
While we see a steep drop in $\chi^{\prime\prime}(T)$ below \TNt{}
in Fig.~\ref{INS_cuts}, instead of a plateau above \TNt{}, as seen
in \BYOO{}, the intensity steadily increases with decreasing temperature
towards a maximum at \TNt{}. We attribute the steady increase predominantly
to the competition between phase 1 and phase 2. Similarly, at very
low temperatures $\chi^{\prime\prime}(T)$ shows a slight upturn,
which we attribute to the remaining ungapped AF1 fraction increasingly
tending towards AF2 order. Therefore, the temperature dependence in
the region $T<T_{\mathrm{N2}}$ is not expected to follow a $\exp(-\Delta/k_{\mathrm{B}}T)$
dependence, but instead is a combination of this reduction in intensity
with the steadily increasing intensity due to the tendency of the
AF1 fraction towards AF2.

The measurements presented have demonstrated that a SOC-induced gap
does emerge in the $3d$-$5d^{3}$ system \SFOO{}, but only below the second
magnetic transition temperature. The neutron diffraction results show
that at all temperatures the AF1 and AF2 magnetic phases are associated
with different structural phases\cite{supp}.  These observations together confirm
the notion that the lattice, magnetic and orbital degrees of freedom
in this material are all intimately linked. It is worth pointing out
that the mixing of Fe and Os sites in this material might
be taken to suggest that there are two phases associated with separated
Fe-rich and Os-rich regions. However, the constant evolution of the
AF1 to AF2 phases below 67 K establishes that there is real competition
between these phases not associated with stoichiometry (although local
stoichiometry may influence \TNt{}). $^{57}$Fe M\"{o}ssbauer spectroscopy
supports this interpretation~\cite{adler_magnetic_2013}. Therefore,
it appears that SOC is an essential component in selection of the
magnetic ground states in Sr$_{2}$FeOsO$_{6}$ as was found for Sr$_{2}$ScOsO$_{6}$\cite{taylor_spin-orbit_2016} suggesting that similar considerations are important for understanding the ground states and high ordering temperatures in other 3d-$5d^3$ TMOs.

\section*{Acknowledgements}

ADC and RSF were supported by the U.S. Department of Energy, Office of Science, Basic Energy Sciences, Materials Sciences and Engineering Division. RM acknowledges support from the Alexander von Humboldt Foundation.  The research at ORNL's Spallation
Neutron Source and High Flux Isotope Reactor was supported by the
Scientific User Facilities Division, Office of Basic Energy Sciences,
U.S. Department of Energy (DOE). Use of the Advanced Photon
Source at Argonne National Laboratory was supported by the U. S. Department
of Energy, Office of Science, Office of Basic Energy Sciences, under
Contract No. DE-AC02-06CH11357. This research was supported in part
by the Center for Emergent Materials an National Science Foundation
(NSF) Materials Research Science and Engineering Center (DMR-1420451).

\bibliographystyle{apsrev4-1}
%

\end{document}